\documentclass{article}
\usepackage{graphicx} 
\usepackage{bm,amsmath,color,dsfont, amssymb}
\usepackage[titletoc,toc,title]{appendix}
\usepackage{graphicx}
\usepackage[section]{placeins}
\usepackage{esint}
\usepackage[hidelinks]{hyperref}
\usepackage[justification=centering]{caption}
\usepackage{cleveref}
\usepackage{subcaption}
\usepackage{float}
\usepackage{multicol}
\usepackage{array} 
\usepackage{authblk}
\usepackage{tabularx}
\usepackage{algorithm, algorithmic}
\usepackage{tgtermes} 
\usepackage{relsize}
\usepackage{wrapfig}
\usepackage{placeins}
\usepackage{siunitx}
\usepackage{bm}
\usepackage{xcolor}
\usepackage[
backend=biber,
style=nature,
]{biblatex}
\usepackage{booktabs}
\usepackage{multirow}

\newcommand{\eps}{\varepsilon}
\newcommand\red[1]{\textcolor{red}{#1}}

\DeclareMathOperator{\EX}{\mathbb{E}}

\begin{document}

\title{Generating realistic metaorders from public data}
\author[1,2,3]{Guillaume Maitrier}
\author[3]{Grégoire Loeper}
\author[4, 2, 5]{Jean-Philippe Bouchaud}

\affil[1]{\textit{LadHyX UMR CNRS 7646, École polytechnique, 91128 Palaiseau, France}}
\affil[2]{\textit{Chair of Econophysics and Complex Systems, École polytechnique, 91128 Palaiseau, France}}
\affil[3]{
\textit{BNP Paribas Global Markets, 20 Boulevard des Italiens, 75009 Paris, France}}
\affil[4]{\textit{Capital Fund Management, 23-25 Rue de l'Université, 75007 Paris, France}
}
\affil[5]{\textit{Académie des Sciences, Paris 75006, France}}

\maketitle

\begin{abstract}
    This paper introduces a novel algorithm for generating realistic metaorders from public trade data, addressing a longstanding challenge in price impact research that has traditionally relied on proprietary datasets. Our method effectively recovers all established stylized facts of metaorders impact, such as the Square Root Law, the concave profile during metaorder execution, and the post-execution decay. This algorithm not only overcomes the dependence on proprietary data, a major barrier to research reproducibility, but also enables the creation of larger and more robust datasets that may increase the quality of empirical studies. Our findings strongly suggest that average realized short-term price impact is not due to information revelation (as in the Kyle framework) but has a mechanical origin which could explain the universality of the Square Root Law. 
\end{abstract}
\tableofcontents

\section{Introduction}
One of the most universal facts in market microstructure is the so-called Square Root Law (SQL), which states that the average price change induced by the execution of a large ``metaorder'' of volume $Q$, executed through a sequence of ``child orders'', is not linear in $Q$, as the classical Kyle theory would predict \cite{kyle1985continuous}, but rather proportional to $\sqrt{Q}$. This surprising effect has been thoroughly documented over the past 40 years and holds for a large variety of markets, trading styles, etc. -- see e.g. \cite{loeb1983trading, grinold2000active, almgren2005direct, toth2011anomalous, bacry2015market, donier2015million, toth2016square, bucci2018slow, AQR, sato2024does} and \cite{bouchaud2018trades, webster2023handbook} for reviews. 

Such studies are chiefly based on proprietary datasets and furthermore reveal other fascinating stylized facts: 
\begin{enumerate}
    \item the SQL is in a first approximation {\it independent} of the time $T$ needed to execute the metaorder, and only depends on the volatility of the asset and the fraction of the total traded volume captured by $Q$ \cite{bouchaud2018trades, webster2023handbook};
    \item the SQL also holds ``inside'' each metaorder: the average price profile is itself a square-root as a function of the currently executed volume. This means that the last child orders impact less than the first ones \cite{maitrier2025double, donier2015million, said2017market};
    \item the square-root impact decays post-execution over the time scale $T$ of the metaorder itself \cite{moro2009market, farmer2013efficiency, bacry2015market, said2017market, bucci2018slow}, with a sharp decay at first and a very slow decay at long times -- with perhaps a small but non-zero permanent component \cite{bucci2018slow, bouchaud2022inelastic}.
\end{enumerate}

These empirical results are of primary importance for both academics and practitioners. The SQL indeed predicts that impact costs are extremely high even for small volumes $Q$, because of the infinite slope of the square-root function at the origin. Neglecting such costs can easily turn a profitable strategy on paper into a money losing machine once implemented, see e.g. \cite{hey2023costmisspecifyingpriceimpact}. From an academic point of view, the explanation of such a square-root dependence is far from trivial -- is it due to information revelation, as many standard economic theories postulate \cite{kyle1985continuous, hasbrouck2007empirical, gabaix2006, farmer2013efficiency, saddier2024bayesian, durin2023squarerootlawsmarket}, or mostly ``mechanical'', as postulated by ``latent liquidity'' theories \cite{toth2011anomalous, mastromatteo2014,donier2015fullyconsistentminimalmodel, donier2016walras, bouchaud2022inelastic}, see also \cite{bouchaud2018trades}.

Despite its significance, empirical research on market impact, specifically when it comes to ``metaorders'', often faces limitations due to data access constraints. Indeed, to track the impact of those metaorders, one should access proprietary datasets, typically held by private institutions, limiting the scope and reproducibility of academic research. Furthermore, such proprietary datasets are often not very large and possibly biased by the trading style of the managers: as emphasized in \cite{bouchaud2018trades, webster2023handbook} market impact and short term trading signals can be difficult to disentangle -- in fact, mainstream economists would claim that ``impact'' is nothing but the correlation between the sign of informed trades and the subsequent price change \cite{bouchaud2010impact}.   

Very recently, Sato and Kanazawa \cite{sato2024does} have been able to access the records of all trades of the Tokyo Stock Exchange (TSE), with (anonymized) trader labels that allowed them to reconstruct all metaorders unambiguously. Their analysis allowed them to confirm once again the validity of the SQL with great precision, and to establish that such a law holds for all stocks individually, when some theories based on the volume distribution or on the autocorrelation of the sign of trades would have predicted systematic deviations from a square-root law \cite{gabaix2006, farmer2013efficiency}. The same dataset has also been used to unveil further, more subtle properties of the SQL \cite{maitrier2025double}.

Such a unique dataset is however, quite unfortunately, inaccessible for open academic research. There have thus been many attempts to create proxies of metaorder impact using the public tape, i.e. the list of all buy and sell market orders executed on lit markets, but without any tags allowing one to track individual traders. To the best of our knowledge, these attempts have been unsuccessful. Identifying metaorder impact with the correlation between order imbalance in a time interval $T$ is clearly completely wrong -- impact is linear for small imbalance and saturates for large imbalance \cite{patzelt2018universal}. It is all but impossible to accurately identify real metaorders within the order flow \cite{tsaknaki2024online}. It is also extremely difficult to generate data that recreate all the stylized facts mentioned above using VAR models or propagator models calibrated on real data, see e.g. \cite{varmodel,naviglio2025estimationmetaorderimpactpublic}. In recent years, machine learning has become a widely used tool for generating limit order book and understand impact, see \cite{nagy2023generative, coletta2022learning}. Nevertheless, all models still struggle to fully capture the characteristics of metaorder execution \cite{cont2023limit}.

To address such challenges, we propose in this paper an algorithm that uses public trade data to generate synthetic, metaorders which lead to price impact indistinguishable from that observed using proprietary datasets. Our method not only circumvents the need for proprietary data but also facilitates the creation of larger and more robust datasets. By adequately aggregating public trade data, we demonstrate that the resulting synthetic metaorders preserve all major characteristics found for real metaorders (SQL, concave execution profiles, post-execution decay), thus providing a valuable tool for both academics and practitioners. 

Our paper is mostly algorithmic and empirical in nature, as we explain our procedure in a clear and reproducible way, and present a sample of the results we have obtained that fully validate our proposal. The theoretical justification for the success of our procedure is not yet completely understood and we will propose a modelling framework in a subsequent publication \cite{ustocome}. But we believe that the fundamental idea is the following: since the SQL is measured for {\it all} metaorders, independently of the trading firm, and since market orders executed in markets mostly originate from such metaorders and are anonymous, the emergence of the SQL cannot heavily rely on the precise matching between market orders and metaorders. This is indeed what was observed in our previous paper \cite{maitrier2025double} using the detailed TSE data as a validation, and that we generalize in the present paper, which is structured as follows: 
\begin{itemize}
    \item Section \ref{sec:Algo} provides a detailed explanation of our synthetic metaorder generation algorithm, with an emphasis on the significance of what we call the ``mapping'' function.
    \item Section \ref{sec:Empirical} presents several evidences of the method's effectiveness in replicating and validating the well-documented empirical facts about metaorders.
\end{itemize}

\section{The Algorithm}\label{sec:Algo}

In this section, we present our algorithm designed to generate random metaorders from publicly available data, which lead to impact properties indistinguishable from actual proprietary data. We define a metaorder as a sequence of trades of the same sign initiated by a given trader within the same trading session. In order to generate {\it synthetic} metaorders, we propose the following algorithm, which requires only public trade data for any asset class (stock, futurs, options etc...). Although using aggregated order book data across multiple venues yields similar results, we recommend using data from a single exchange (ex : Euronext, Nasdaq, CME etc...).

\begin{algorithm}[H]
\caption{Generating Synthetic Metaorders}
\begin{algorithmic}[1]
\STATE Load and clean trade data for given stock and date (ex : remove opening and closing periods)
\STATE Compute daily traded volume $V_D$ and intraday volatility $\sigma_D$, which we define for simplicity as:
\begin{equation}
    V_D = \sum_{i} q_i, \quad \sigma_D = \frac{\max(p_t) - \min(p_t)}{p_0}
\end{equation}
\STATE Randomly assign trades to traders using a {\it mapping function} while preserving the true chronological order of trades
\STATE Sort trades by traders and timestamp.
\STATE Define a metaorder as  a sequence of trades of same signs from the same trader.
\STATE Compute metaorder features:
\begin{itemize}
    \small
    \item Log price at metaorder start and end.
    \item Number of child orders in metaorder.
    \item Volume traded within the metaorder.
    \item Any other quantities that may be relevant.
\end{itemize}
\STATE Aggregate metaorder statistics and return only those with more than one child order.
\end{algorithmic}
\end{algorithm}

\subsection*{The mapping function}
The important feature of the algorithm is the mapping function, which assigns synthetic trader IDs to each market order executed on a particular day. This function is crucial: if one possesses a proprietary dataset \cite{toth2011anomalous} or an exhaustive one such as the TSE dataset \cite{sato2024does}, one knows this mapping exactly at least for a given set of market orders, and this allows one to measure the square root law in the usual manner, namely (see \cite{bouchaud2018trades})
\begin{equation}\label{eq:SQL}
    \frac{I(Q)}{\sigma_D} = Y\sqrt{\frac{Q}{V_D}}, \quad \text{with } \begin{cases}
        & I(Q) = \EX[\eps \cdot (p_e-p_s)] \\
        & Y \in [0.5, 1],
    \end{cases} 
\end{equation} 
where $p_s$ is the start mid price, just before the execution of the first child order and $p_e$ is the end mid price, just before the execution of the market order immediately following the last child order. Now, as highlighted in \cite{maitrier2025double}, the introduction of random variations in the matching between real traders and orders, the square impact law is preserved, including its prefactor $Y$. Here, we show that even for a mapping function totally agnostic of the true mapping, we still recover the correct impact. \\
\par
We gave this mapping function only two degrees of freedom: the number of different traders in a given day and the distribution of their trading frequency, that is, the fraction of orders they participate to. We will show later that the impact law is only weakly dependent on those parameters, as expected. However, these parameters directly influence both the number and average length of the generated random metaorders. Therefore, considering the stock's liquidity (i.e. the number of trades per day and the average volume per trades), it may be necessary to set these parameters within an appropriate range to generate coherent metaorders.\\

Below is the pseudo-code for the mapping function:

\begin{algorithm}
\caption{Mapping Function}
\begin{algorithmic}[1]
    \STATE \textbf{Initialization:} Let $N$ be the number of traders, and $F$ a probability law 
    \STATE Generate $f_i \sim F$  for $i=1,\dots,N$.
    \STATE Define $p_i = \frac{f_i}{\sum f_i}$ for each agent.
    \STATE Compute cumulative probabilities $c_i = \sum_{j=1}^{i} p_j$, $c_0=0$.
    \FOR{each order in the market}
        \STATE Draw a random variable $U \sim U(0,1)$.
        \STATE Find the trader $i$ such that $c_{i-1} \leq U < c_i$.
        \STATE Assign the order to agent $i$.
    \ENDFOR
\end{algorithmic}
\end{algorithm}

We have evaluated the robustness of our algorithm using different values of $N$ and two types of trader frequency $f$ distributions: a power-law $P(f) \propto f^{-\alpha}$ and a homogeneous $f \equiv f_0$. In real markets, the distribution of trader participation is well approximated by a power-law distribution, where a small number of traders account for a significant portion of executed orders, while most traders participate less frequently, see \cite{maitrier2025double}. \\

{Note that a key aspect of this mapping function is that it corresponds to sampling without replacement. We find this feature essential for recovering the SQL.\\}

Now the procedure is established, we focus next on empirical results and demonstrate the robustness of our method with which we can generate an unlimited number of realistic metaorders.  

\section{Recovering Metaorder Stylized Facts}\label{sec:Empirical}

In this section we report the results of our empirical investigations using synthetic metaorders. A series of sanity checks have been performed to rule out any trivial artifacts (see Appendix). For example we check that by randomly flipping the sign of market orders we measure zero impact, as it should be -- impact is {\it not} merely related to volatility \cite{bucci_impact_vol2019}. 

\subsection{Peak impact: the Square Root Law}

\begin{figure}[H]
    \centering
    \includegraphics[width=0.6\linewidth]{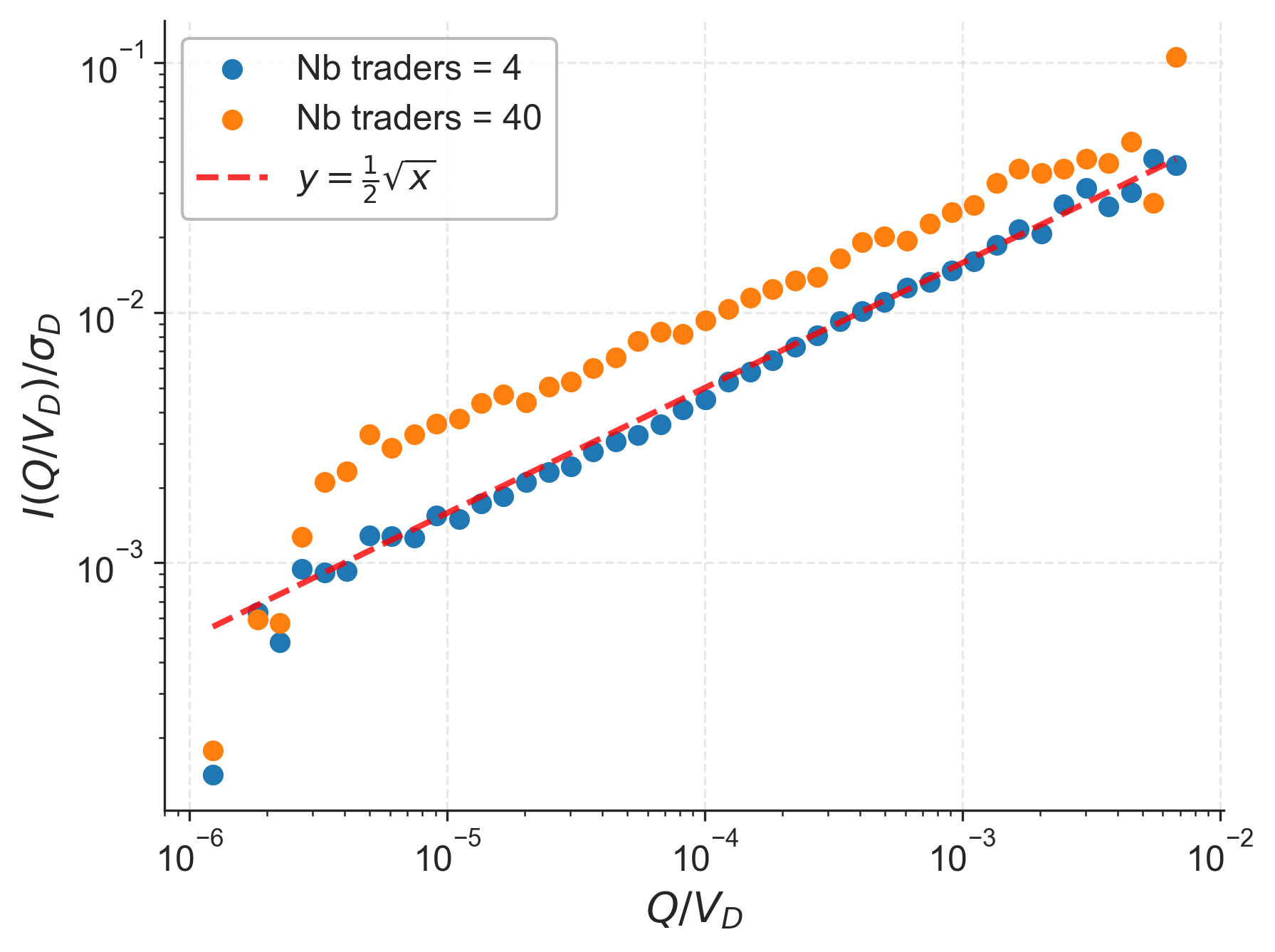}
    \caption{Retrieving the Square Root Law for futures on the Eurostoxx, trade data from 2016 to 2018. We used a mapping function with 4 trades (resp. 40 trades) for the blue curve (resp. orange), and homogeneous distribution of their trading frequencies, resulting in approximately 3 million metaorders in both cases. We see that, by fine-tuning the number of traders, one may recover the correct prefactor $Y\approx 0.5$}
    \label{fig:SQL_Eurostoxx}
\end{figure}

We first tested our algorithm on a very liquid asset: the EuroStoxx futures contract from September 2016 to August 2018. We used public trade data, selecting only the Eurex exchange. We obtain a remarkably clean square root law over four decades, see Figure \ref{fig:SQL_Eurostoxx}, with a noisy region for very small $Q/V_D \leq 5\times10^{-6}$, which might possibly be considered as a linear, as in \cite{Buccicrossover}. Note that we do not only recover the square-root dependence on $Q$ but also the correct {a realistic} prefactor in Eq. \eqref{eq:SQL}, with $Y= 0.5$, {as found in \cite{toth2011anomalous,sato2024does}}.

\begin{figure}[H]
    \centering
    \begin{minipage}{0.49\linewidth}
        \centering
        \includegraphics[width=\linewidth]{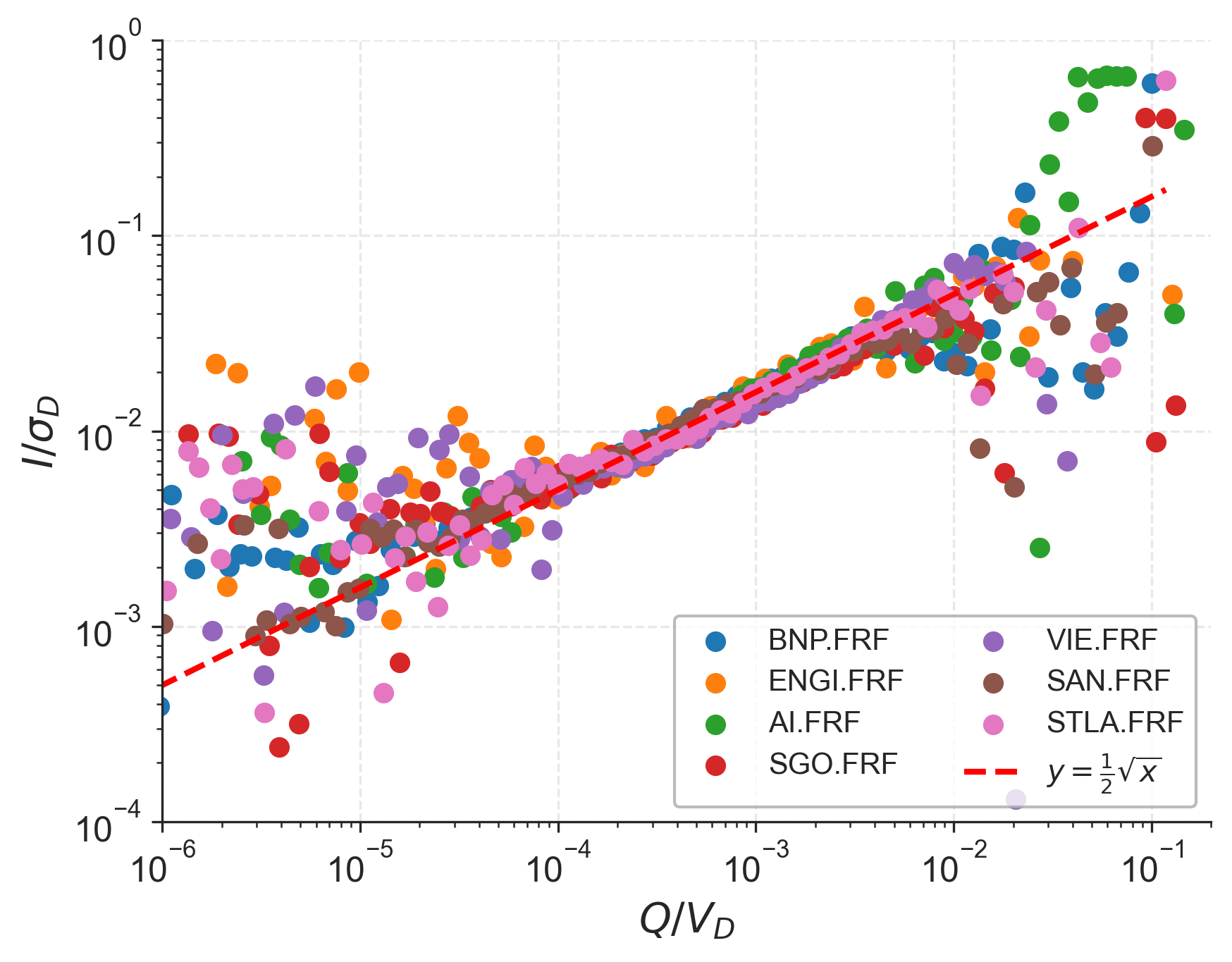}
    \end{minipage}
    \begin{minipage}{0.49\linewidth}
        \centering
        \includegraphics[width=\linewidth]{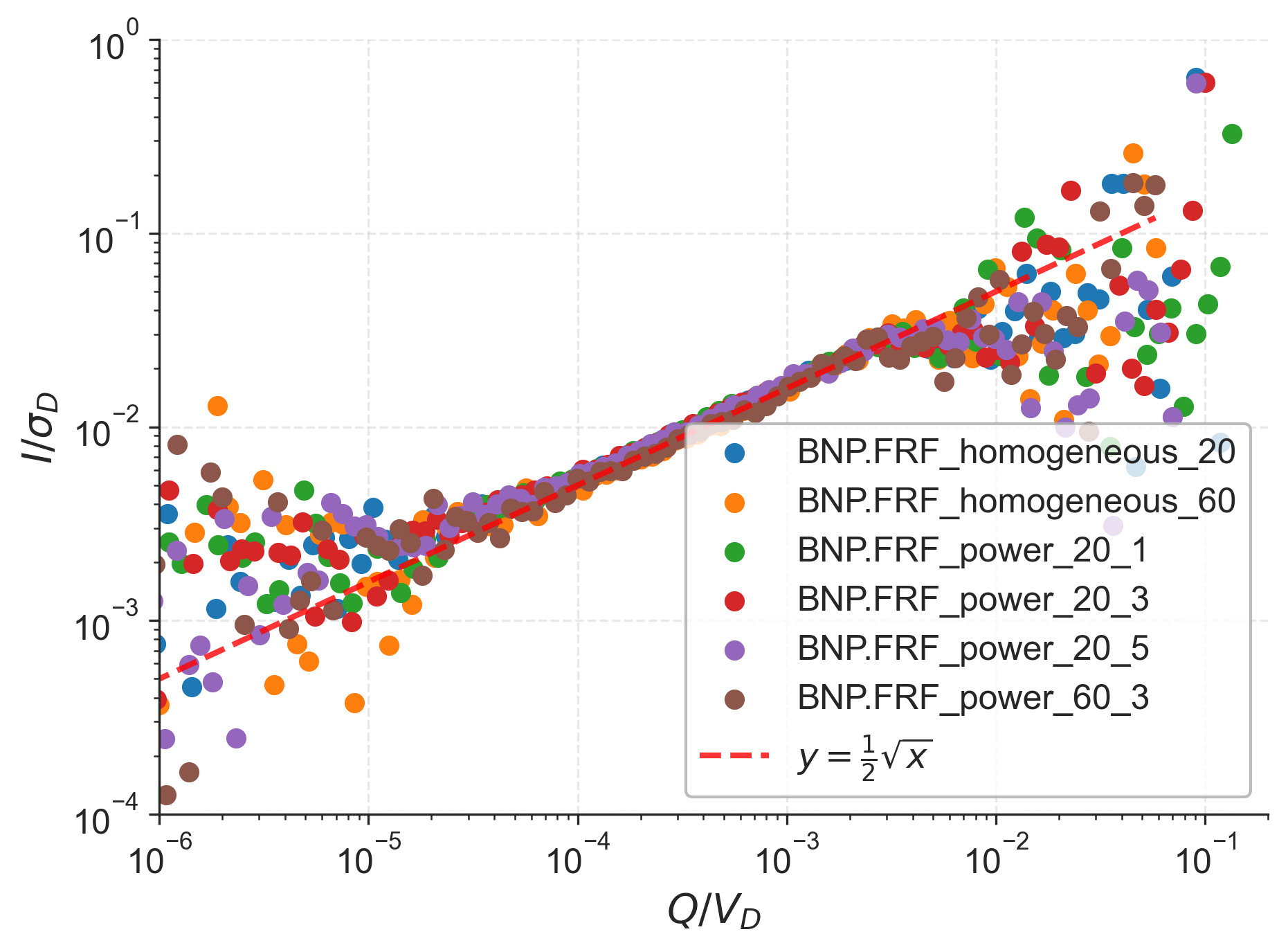} 
    \end{minipage}
    \caption{\textbf{Left:} Retrieving the Square Root Law for various stocks traded on the Paris Stock Exchange. Synthetic metaorders were generated using our algorithm, specifying 20 traders and a homogeneous distribution of their trading frequency. Stocks were traded on Euronext between 2021 and 2023. \textbf{Right:} Verifying the robustness of the algorithm in respect to variations in the mapping function parameters. Legend represents (Stock Name; Type of Distribution; Number of traders; Power law exponent). Data from Paris Stock Exchange, between 2021 and 2023.}
    \label{fig:SQL_French}
\end{figure}

We also tested these results on single stocks by selecting a basket of seven stocks traded on the Paris Stock Exchange, between January 2021 and December 2023. We used our algorithm to generate approximately 3 millions of random metaorders per stock, imposing 20 traders and a homogeneous trading frequency distribution. Once again, we obtained very precise results, with almost no variation in the prefactor ($Y \approx 0.5$), see Figure \ref{fig:SQL_French}. This could be explained by the fact that selected stocks are among the most liquid ones traded on the PSE, and thus may be quite similar. We also tested for a specific stock (BNP Paribas) the dependence of the SQL on the input parameters, i.e. the number of traders and the distribution of their trading frequencies. Again, we found no significant variations in the impact function. However, we do acknowledge that for some assets, particularly illiquid ones, one may have to fine-tune those parameters to recover the SQL. It is necessary to be statistically close enough to the real mapping function. The same remark also applies to the Y-ratio. While we mostly show impact functions with a realistic prefactor, i.e., close to 0.5, we have also encountered stocks where the prefactor was slightly higher of lower, but typically of order one, see Figure \ref{fig:SQL_Eurostoxx}.

\subsection{Role of metaorder duration}

One of the major enigmas of metaorder price impact is that, contrary to a priori expectations, the impact remains independent of the metaorder duration $T$. This phenomenon is a natural consequence of the SQL as written in Eq. \eqref{eq:SQL}: indeed as $T$ is varied the volatility contribution scales as $\sqrt{T}$ whereas the total traded volume scales as $T$, which means that the explicit $T$ dependence cancels from $I(Q)$. This property was extensively studied in \cite{bucci_impact_vol2019} and confirmed in \cite{maitrier2025double} both for real and synthetic metaorders based on the TSE dataset. We show in Fig. \ref{fig:enter-label} that this property also holds for our synthetic metaorders.  

\begin{figure}[H]
    \centering
    \includegraphics[width=0.6\linewidth]{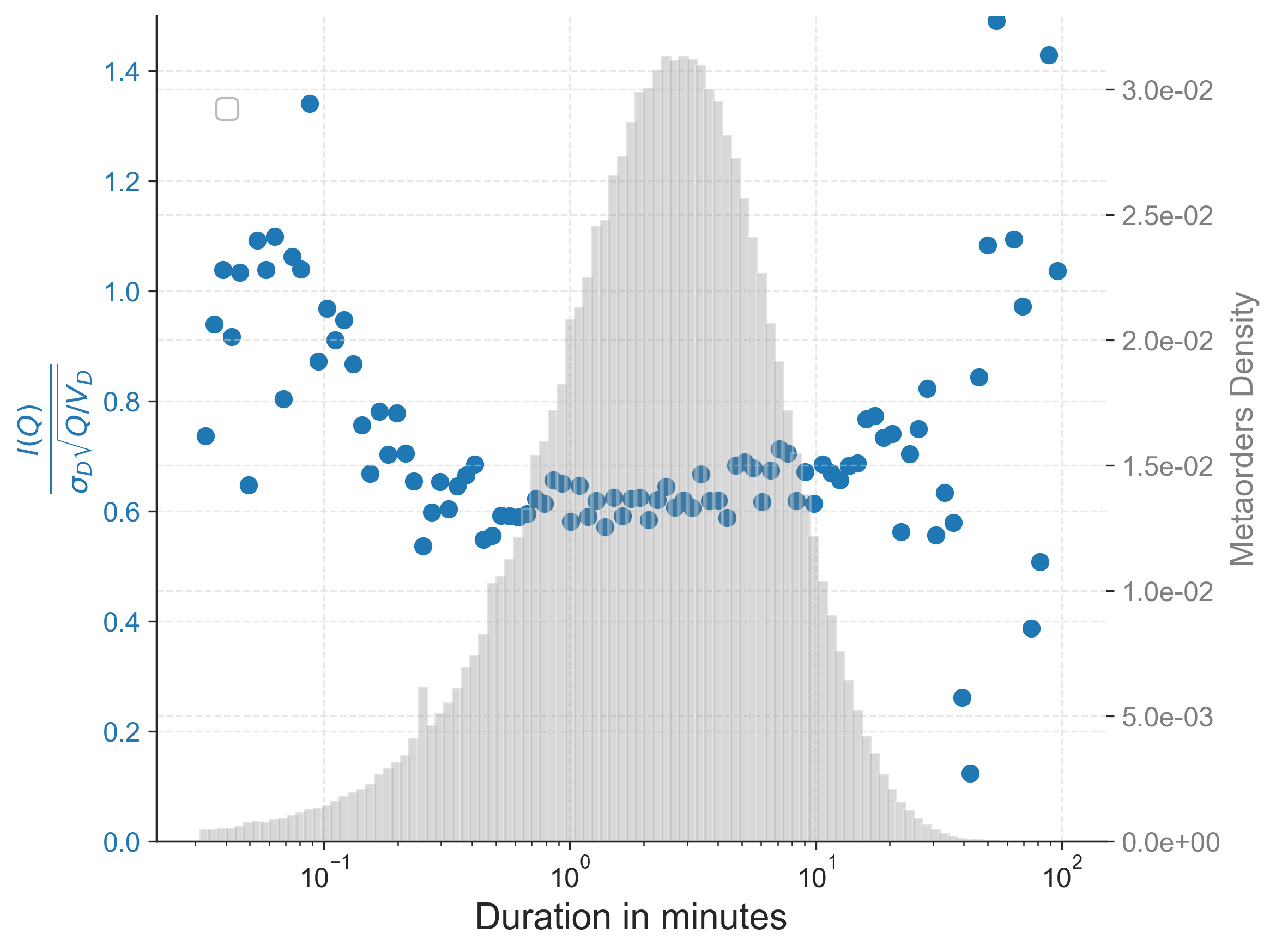}
    \caption{Approximate independence of metaorder impact $I(Q)$ with respect to the metaorder duration $T$ (expressed in minutes). The blue dots represent the impact divided by the volume contribution as a function of duration, which is approximately constant for $T \gtrsim 30$ seconds, and on average equal to $Y=0.6$. The grey histogram shows the distribution of metaorders durations in minutes. Synthetic  metaorders were generated with BNP Paribas share price between 2020 and 2023, for 20 homogeneous traders.}
    \label{fig:sql_duration}
\end{figure}

{Figure \ref{fig:sql_duration} also reveals that, on average, synthetic metaorders constructed using our method are shorter than those typically found in proprietary datasets. For instance, it is not uncommon for firms like CFM to execute metaorders over an entire trading day. However, the durations of our synthetic metaorders are consistent with the average metaorder duration observed in the TSE dataset, which is typically also around 2-5 minutes, see \cite{maitrier2025double}. In any case, one can adjust metaorder duration by tuning the mapping function's parameters. 

\subsection{Concave profile during metaorder execution}

Beyond the peak impact $I(Q)$, our method is also particularly effective in reproducing other stylized facts, such as the concave profile during metaorder execution, see \cite{moro2009market,bacry2015market, said2017market, donier2015million}. Indeed, as proposed in \cite{bouchaud2018trades}, the average price impact during the execution of the metaorder reads: 
\begin{equation}
    \mathcal{I}(\phi Q) = \sqrt{\phi}I(Q), 
\end{equation}
where $\phi \in [0,1]$ is the fraction of executed volume, with $\phi=0$ at the start of the metaorder and $\phi=1$ at the end. 
\begin{figure}[H]
    \centering
    \includegraphics[width=0.5\linewidth]{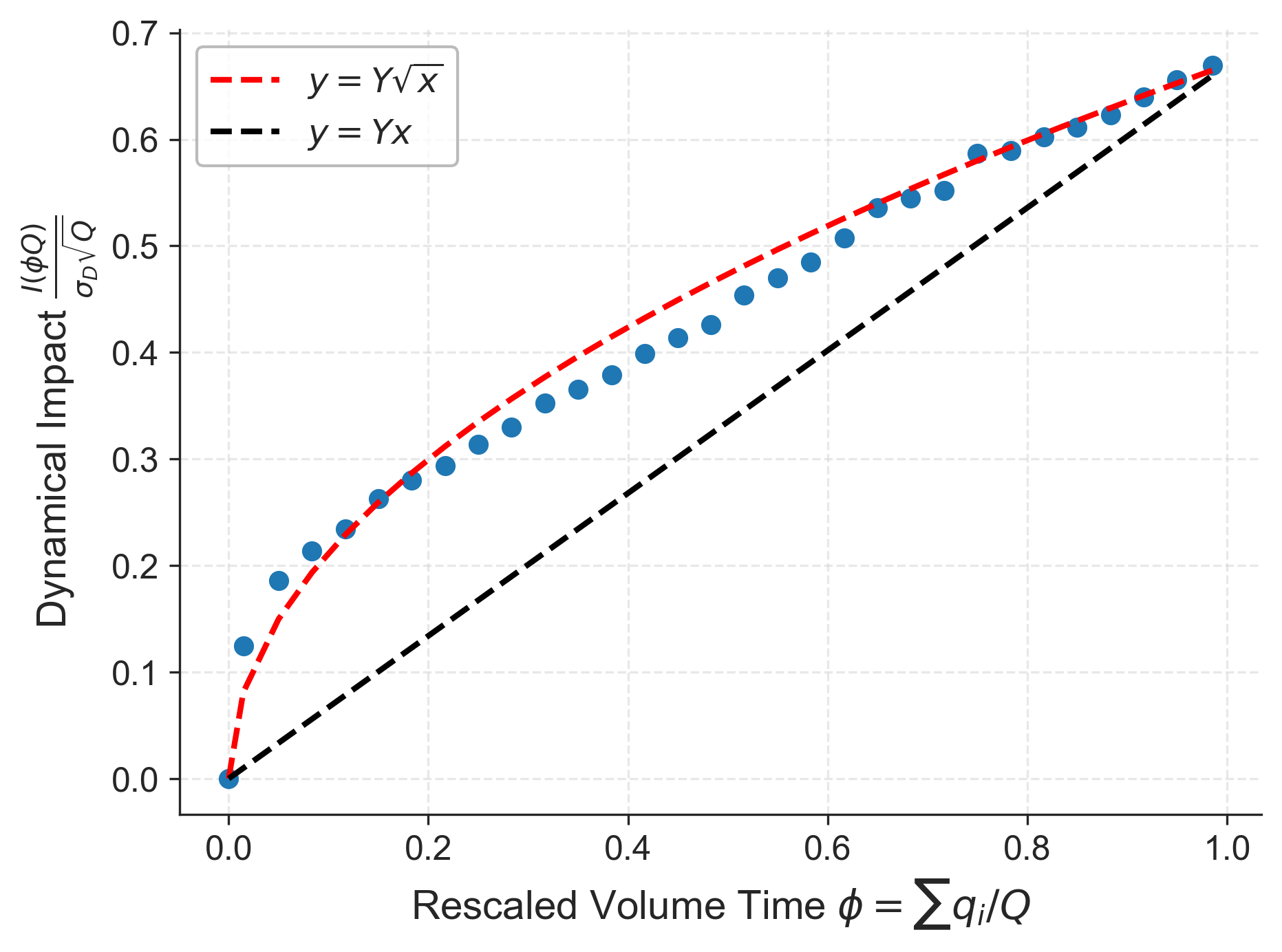}
    \caption{Concave profile during metaorder execution. Random metaorders constructed on BNP Paribas stock price, data from 2020 to 2023, with mapping function parameters of 20 traders and a power law distribution for their trading frequencies with an exponent of $\alpha = 2$. We selected only metaorders having more than 5 child orders.}
    \label{fig:profile_execution}
\end{figure}
Such a concave profile can be explained within the latent liquidity framework, which predicts that the latent limit order book is locally linear (LLOB) \cite{donier2015fullyconsistentminimalmodel,bouchaud2018trades}. The fact that this profile also holds for synthetic metaorders is another indication that, on average, the latent order book is indeed present and provides liquidity for all metaorders. This could be related to the action of market makers, as proposed in \cite{maitrier2025double}. That said, the concavity is in fact crucial for market efficiency : it may be a key element to ensure the diffusivity of prices as argued recently in \cite{sato2025exactlysolvablemodelsquareroot}. However, we believe that the framework developed in that paper is insufficient to accurately describe reality, as it lacks a crucial component: metaorder decay, to which we turn next. A unified theory of price impact that incorporate all known ingredients (autocorrelation of the sign of the trades, square-root impact, impact decay) is still under construction, a topic on which we hope to report soon \cite{ustocome}.

\subsection{Metaorder decay post execution}

Impact decay has been subject to controversy, even when it is of crucial importance for optimal execution schedules. Indeed, assuming permanent impact or accounting for impact decay leads to radically different trading policies. What makes the empirical study of this problem particularly difficult is the fact that price variance increases linearly with the time elapsed since the end of execution, leading to large errors in the determination of impact decay. We know that metaorder impact during execution is generally small relative to the volatility (see Eq. \eqref{eq:SQL} for small $Q/V_D$), this predicament is especially strong for metaorder decay: the signal-to-noise ratio significantly worsens when analyzing extended timescales after execution, highlighting the need for large metaorder datasets. 
\\ 
An initial line of research suggested that there is a permanent impact after execution, at approximately $2/3$ of the peak impact, meaning that the price after execution is equal to the average paid during execution, see \cite{farmer2013efficiency}. However, a later empirical study found that upon closer inspection the impact eventually decays to zero over a much longer timescale (several days). This can potentially be mistaken for a permanent impact of about \(2/3\) by the end of the trading day \cite{bucci2018slow}. However, if the decay of the impact is evaluated over multiple days, a clear decay of impact is observed -- although the long term fate is difficult to ascertain and it is plausible that a small permanent impact exists \cite{benzaquen2017marketimpactmultitimescaleliquidity,gabaix2021search,bouchaud2022inelastic}. 
   
Interestingly, using our synthetic metaorders, we precisely replicate the fit observed in \cite{bucci2018slow} with real data. Assuming a propagator decaying as $G(t) \approx t^{-\beta}$ with $\beta < 1$, the rescaled impact after execution for a metaorder of size $Q$ and duration $T$ can be expressed as: 
\begin{equation} 
    \mathcal{I}(Q,z) = {I}(Q)\left(z^{1-\beta} - (z-1)^{1-\beta}\right),    
\end{equation}
with $z = t/T \geq 1$ and $t=0$ corresponds to the start of the metaorder. Such a decay is fast at the beginning (with a sharp singularity $-(t-T)^{1 - \beta}$ and a slow relaxation at long times (as $t^{-\beta}$).

\begin{figure}[H]
    \centering
    \includegraphics[width=0.5\linewidth]{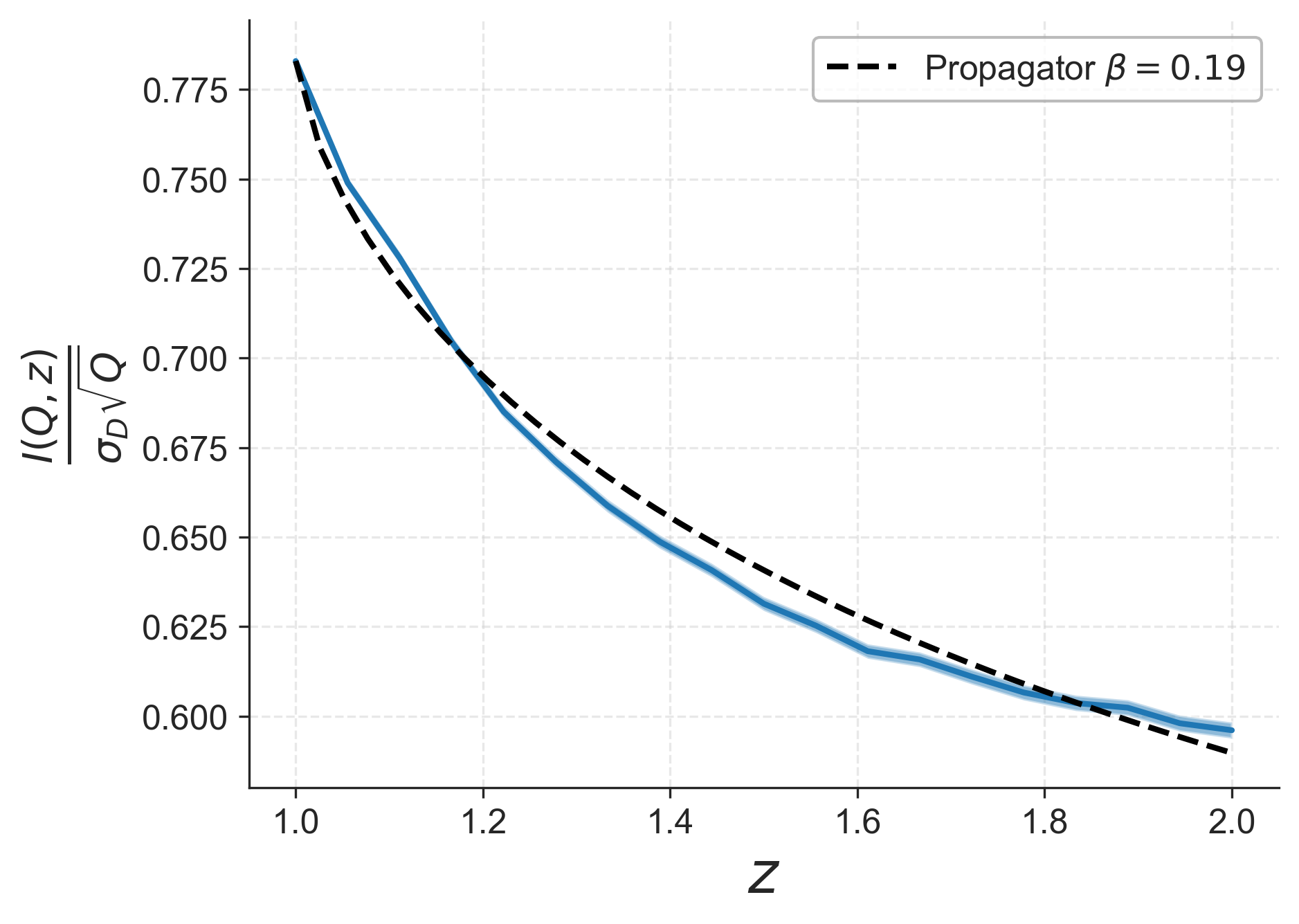}
    \caption{Price relaxation post metaorder execution, well fitted by the impact predicted by the propagator model, with the value $\beta \approx 0.2$. We used synthetic metaorders generated on BNP Paribas share price between 2021 and 2023, using a mapping function of parameters of 4 traders and power law distribution of exponent $\alpha =2$. We kept only metaorders with more than 5 child orders.}
    \label{fig:decay}
\end{figure}
Hence, we confirm using our synthetic metaorder database that impact not only decays after execution, but rather interestingly it decays exactly as found by Bucci et al. \cite{bucci2018slow} using real metaorders, with a value of $\beta=0.2$ very close to the one reported there ($\beta=0.22$). This is also the value predicted by the propagator model $\beta = \frac{1-\gamma}{2}$ \cite{bouchaud2003fluctuations}, where $\gamma$ describes the power-law decay of the autocorrelation of trades, which is typically found to be around 0.5 for most stocks, see \cite{Sato_2023}. 


\section{Conclusion}

In this paper, we introduced a straightforward yet surprisingly effective algorithm for generating realistic metaorders from public trade data. This approach offers a robust solution to a longstanding dataset challenge in price impact research, which has traditionally relied on proprietary data. Using this algorithm, we were able to recover all the salient stylized facts reported in the existing literature, specifically: the Square Root Law and its independence with respect to metaorder duration; the concave profile during metaorder execution; and the slow power-law decay after execution. We also confirm that this decay is effectively captured by the prediction of the propagator model, in line with previous studies. Of course, to generate even more realistic metaorders, some refinements could be made to this algorithm, particularly when it comes to the mapping function. However, our goal was to provide a highly reproducible yet effective algorithm in this area of research, which, by essence, is often not transparent when it comes to data. We therefore believe this could be a valuable tool for both practitioners and academics to enhance the quality of their empirical studies.
\par
On the other hand, it also serves as further evidence that there is complete decorrelation between putative prediction signals (a.k.a. short term alpha) and the square-root law governing the average realized price impact (at least in the short term). By construction, a synthetic metaorder has no connection to a trader’s intention to trade based on a predictive signal. Yet, its impact is indistinguishable from that of a real metaorder, which, by contrast, may be executed by a trader who follows such a signal. Hence, this result supports a mechanical origin of price impact, which may also explain its universal nature. This naturally leads to the next key question for future research: what are the theoretical foundations underlying our findings? Indeed, if price impact is purely mechanical, then one might expect a theory purely based on order flow dynamics, endogenously in a sense, could fully capture this phenomenon. Evidence seems to favour the theory of the latent limit order book \cite{sato2024does}, although some modifications may be required \cite{maitrier2025double}. We aim to come back to this fascinating question in an upcoming paper. Finally, from an empirical perspective, a natural next step would be to extend this method to measure cross-impact, which is even more influenced by the scarcity of datasets than self-impact \cite{hey2023costmisspecifyingpriceimpact}.

\section*{Acknowledgments}
We would like to thank  {Doyne Farmer}, Natascha Hey, Kiyoshi Kanazawa, Fabrizio Lillo, James Ridgway, Yuki Sato and Bence Toth for helpful discussions and suggestions. This research was conducted within the Econophysics \& Complex Systems Research Chair, under the aegis of the Fondation du Risque, the Fondation de l'\'Ecole Polytechnique and Capital Fund Management.

\printbibliography

\appendix

\section*{Appendix: Sanity Checks and Validation of the Algorithm}\label{sec:appendix}

We provide here a non-exhaustive list of sanity checks one may perform easily to validate the coherence of generated metaorders.

\paragraph{Validating the Consistency of Metaorder Size and Duration \\}
We know that the Square Root Law is independent of the specific microscopic parameters of a market, such as the distributions of metaorder size and duration, as discussed in \cite{sato2024does}. Their empirical analysis is particularly important to support the universalism of such theory. However, even though it may not modify the impact function, it may matter to verify that our generated metaorders are statistically coherent, see Figure \ref{fig:appendix_distribution}. 

\begin{figure}[H]
    \centering
    \begin{minipage}{0.49\linewidth}
        \centering
        \includegraphics[width=\linewidth]{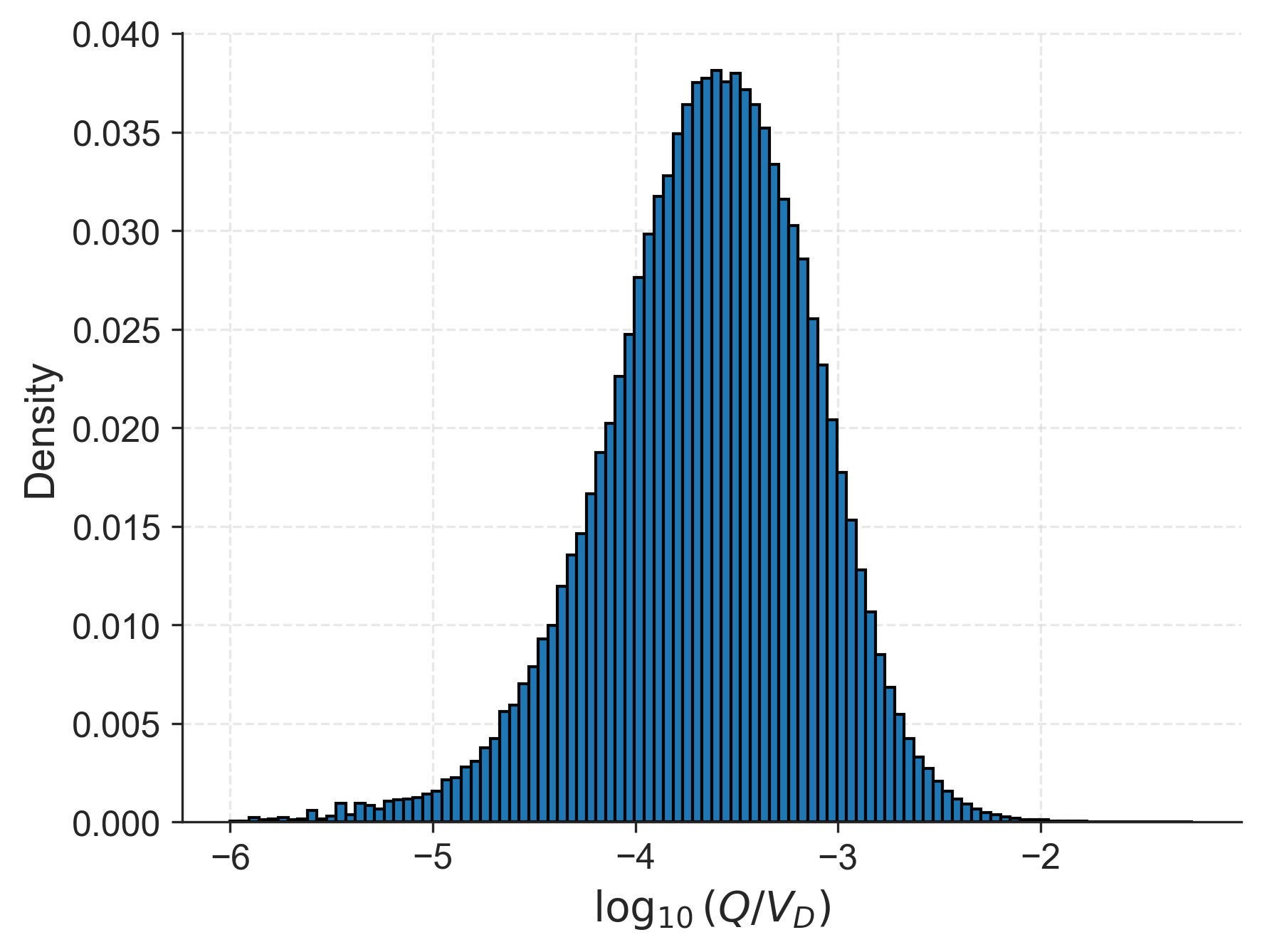}
    \end{minipage}
    \begin{minipage}{0.49\linewidth}
        \centering
        \includegraphics[width=\linewidth]{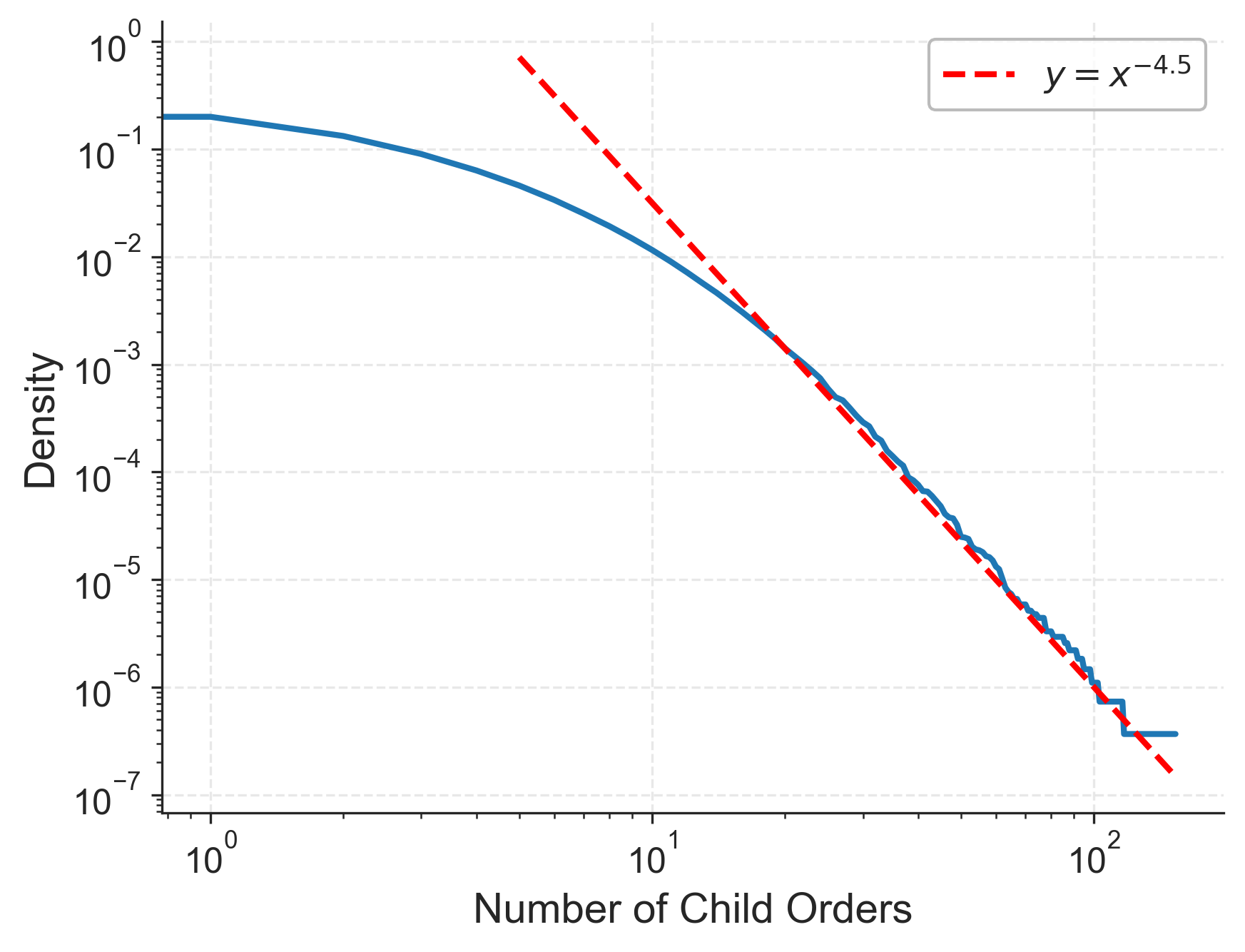} 
    \end{minipage}
    \caption{\textbf{Left:} Distribution of the size of synthetic metaorders, for BNP Paribas, data from 2021 to 2023 generated with a mapping function of parameters 10 traders and homogeneous distribution. \textbf{Right:} On the same dataset, distribution of the number of child orders per metaorders, fitted by a power law of exponent $1 + \mu = 4.5$.}
    \label{fig:appendix_distribution}
\end{figure}

With this algorithm and those parameters, we obtained metaorders distributed around $10^{-3} V_D$, which is typically the order of magnitude one may expect for a metaorder. We also retrieve a power law distribution of the length of metaorders, in line with the Lillo-Mike-Farmer theory \cite{lillo2005theory} and empirical studies, see \cite{sato2023inferring}. However, one may note that the decay is much faster than expected, with $\mu = 3.5$ whereas theory suggests that $1 \leq \mu_t \leq 2$.

\paragraph{Modifying public trade data\\}
To ensure that our algorithm is not merely generating random metaorders with the correct impact due to an unknown construction bias, we conducted several tests that removed information from the public data. One of the simplest consists in randomly shuffling trade signs, referring to buy or sell, in the public data, and then perform again the aggregation method. 

\begin{figure}[H]
    \centering
    \includegraphics[width=0.5\linewidth]{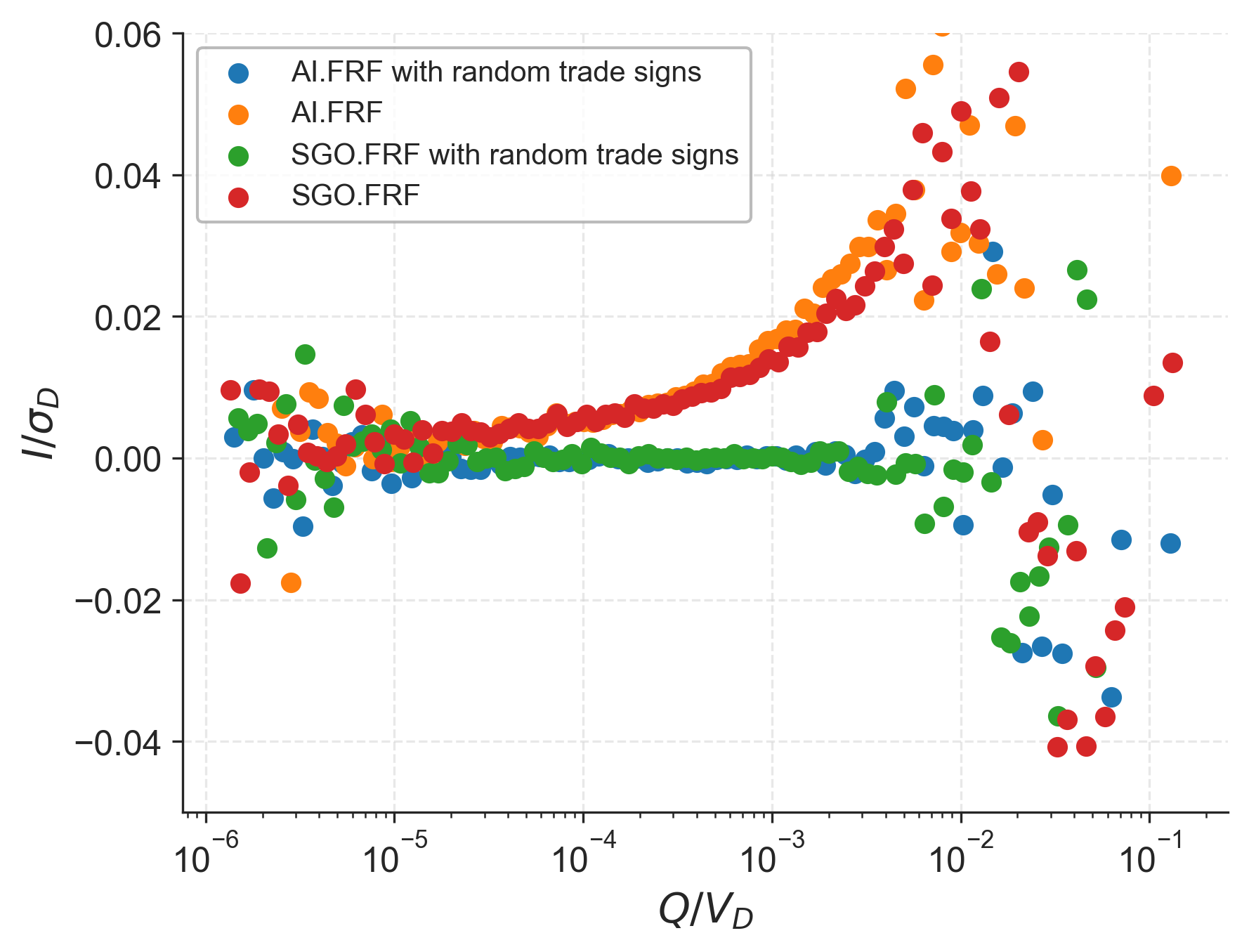}
    \caption{Comparison of the impact law for random metaorders created using public data versus trade data where the signs of market orders have been randomly shuffled. As expected, the average impact is null when the real signs time series is modified. Data from 2020 to 2023, generated using a mapping function with parameters (20 traders, homogeneous distribution).}
    \label{fig:enter-label}
\end{figure}

We also verified that randomly modifying the chronology of public trades execution, keeping real permanent impact for each trade affects the impact law. More generally, to obtain the correct impact, it is important to keep the original trade data as it is, indicating that some information is indeed present. Our argument is that this information is not associated with economic considerations but rather with flow reactions and liquidity responses.

\paragraph{Synthetic metaorders constructed on synthetic prices\\}

A natural question arising from this empirical study is whether public trade data are  actually needed to construct synthetic metaorders that verify the SQL. In other words, is there a specific element or piece of information in real trade data that distinguishes actual prices from random prices? 

A possible experiment is to generate a basic synthetic price. This can be achieved by first creating a sequence of random volumes and random signs. Each order is then assigned a signed instantaneous impact, which can be a linear or concave function of the volume. Then, returns read : 
\begin{equation}
    r_t = \eps_t q^\chi, \quad \chi \in \{0,\frac{1}{2},1\}, \quad q\sim \mathcal{U}(q_{min},q_{max})
\end{equation}
In the following simulation we set $q_{min} = 1$ and $q_{max}= 100$, but the same results hold for $ q = q_{\min} = q_{\max} = 1$, just generating synthetic metaorders with a lower average size and variance. Based on \cite{maitrier2025double}, we fixed $\chi = 0.5$.

This process produces synthetic trade data that can serve as a foundation for constructing random metaorders. However, metaorders constructed with this algorithm won't exhibit SQL, but a linear impact function, see Figure \ref{fig:linear_impact}.

Since trade signs are known to be autocorrelated, it is tempting to leverage this well-established stylized fact to recover the square root law. We generate a series of trade signs with an autocorrelation $C(l) = \langle \eps_t \eps_{t+l} \rangle = l^{-\gamma}$ and we naturally impose a propagator-type impact $G(t) \approx t^{-\beta}, \quad \beta = \frac{1-\gamma}{2}$, see \cite{bouchaud2018trades}, to recover price diffusivity. But again, by doing so, one will still obtain a linear impact for synthetic metaorders, see Figure \ref{fig:linear_impact}.

\begin{figure}[H]
    \centering
    \includegraphics[width=0.6\linewidth]{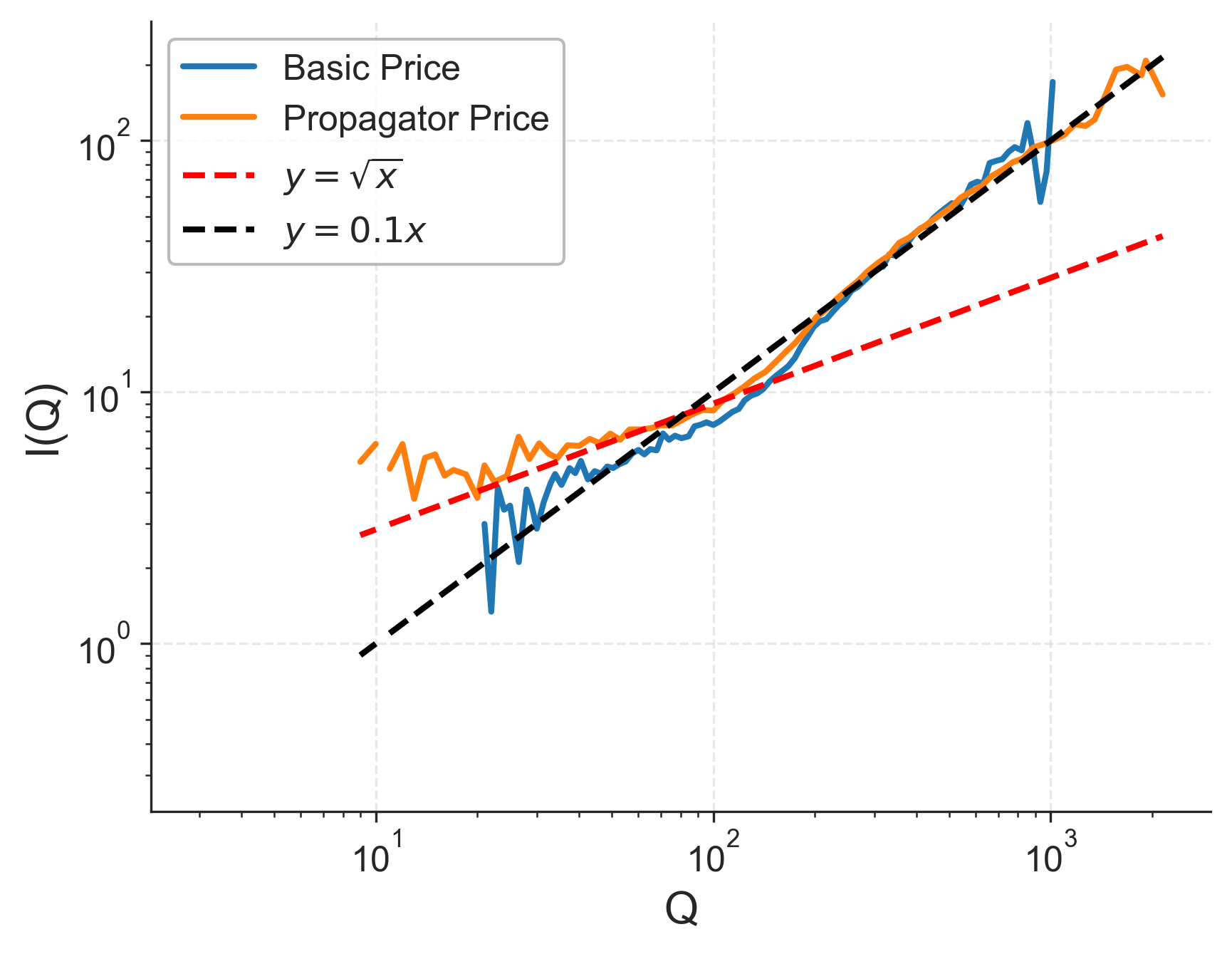}
    \caption{Retrieving a linear impact when constructing synthetic metaorders on a synthetic price. The basic price is built using uniformly distributed volumes ($q_{min} = 1$ and $q_{max} = 100$ and a series of uncorrelated random signs. The instantaneous impact reads $r_t = \eps_t q^\chi$ with $\chi = 0.5$. Synthetic metaorders were constructed with 4 traders and a power law distribution of exponent $\chi = 2$.The propagator price refers to a synthetic price constructed with autocorrelated signs,  $C(l) = \langle \eps_t \eps_{t+l} \rangle = l^{-\gamma}$ and a transient impact that decays according to the propagator model : $G(l) = t^{-\beta}$}
    \label{fig:linear_impact}
\end{figure}

\end{document}